\documentclass[iop]{emulateapj}
\usepackage{amsmath}

\newcommand{\re}{$R_{\oplus}$}

\usepackage{color,hyperref}
\newcommand{\kepler}{\textit{Kepler}}

\shorttitle{Hot Jupiters to Super-Earths}
\shortauthors{Valsecchi, Rasio, \& Steffen}

\begin{document}

\title{From Hot Jupiters to Super-Earths via Roche Lobe Overflow}

\author{Francesca Valsecchi, Frederic A.\ Rasio, \& Jason H.\ Steffen}
\affil{Center for Interdisciplinary Exploration and Research in
  Astrophysics (CIERA)} 
\affil{Department of Physics and Astronomy,
  Northwestern University} 

\begin{abstract}
Through tidal dissipation in a slowly spinning host star the orbits of many hot Jupiters may decay down to the Roche limit.
We expect that in most cases the ensuing mass transfer will be stable.
Using detailed numerical calculations 
we find that this evolution is quite rapid, potentially leading to complete removal of the
gaseous envelope in a few Gyr, and leaving behind an exposed rocky core (``hot super-Earth''). Final orbital periods are quite sensitive 
to the details of the planet's mass--radius relation, 
and to the effects of irradiation and photo-evaporation, but could be as short as a few hours, or as long as several days.
Our scenario predicts the existence
of planets with intermediate masses (``hot Neptunes'') that should be found precisely at their Roche limit and in the process of
losing mass through Roche lobe overflow. The observed excess of small single-planet candidate systems observed by \kepler\ may also be 
the result of this process.  If so, the properties of their host stars should track those of the hot Jupiters. 
Moreover, the number of systems that produced hot Jupiters could be 2--3 times larger than one would infer from contemporary observations.
\end{abstract}
\keywords{Planetary Systems: planet-star interactions--planets and satellites: gaseous planets--stars: evolution--stars: general--(stars:) planetary systems}

\section{Introduction} \label{Intro}
Our current understanding of tidal dissipation in solar-like stars suggests that the orbits of the shortest-period hot Jupiters are rapidly decaying, driven by the Darwin instability (e.g., Rasio et al.\ 1996). How these giant planets reached their current orbits is still a matter of debate. In \cite{ValsecchiR+14} (hereafter VR14) 
we demonstrated that, including the effects of inertial wave dissipation \citep{Lai12}, stellar tides can account for the observed distribution of misalignments between the stellar spin and the planetary orbital angular momentum (e.g., \citealt{WinnFAJ10,Albrecht+12}). Moreover, orbital decay naturally explains the presence of hot Jupiters with orbital separations $a$ less than {\it twice\/} the Roche limit separation 2$a_{\rm R}$ (\citealt{Valsecchi+14edge}, hereafter VR14b). These results support a high-eccentricity migration scenario for the formation of hot Jupiters \citep{RasioFord96,WuMurray03, FordRasio06,FabryckyTremaine07,Nagasawa08,Jackson+08,MatsumuraPR2010,WuLithwick11,Naoz+11,PlavchanBilinski13}.

Several previous studies have considered the fate of giant planets that reach $a_{\rm R}$. 
\cite{TrillingBGLHB1998} investigated stable mass transfer (hereafter MT) from a 
giant planet to its stellar host when the latter is still young and rapidly spinning, as one way of halting disk migration and producing a hot Jupiter.
\cite{ChangGuBodenheimer10} studied the orbital expansion resulting from Roche-lobe overflow (hereafter RLO) of young hot Jupiters inside the magnetospheric cavity of a protoplanetary disk, to explain the absence of low-mass giant planets within $\sim\,$0.03 AU.
Many investigations have simply assumed that, whenever a planet reaches $a_{\rm R}$, it is quickly destroyed and its material accreted by the star
\citep{Jackson+09,Metzger+12,SchlaufmanWinn13,TeitlerKonigl14, ZhangPenev14}.
This could yield a detectable transient signal (e.g., \citealt{Metzger+12}) and, if the star is spun up significantly through accretion \citep{Jackson+09,TeitlerKonigl14,ZhangPenev14}, it could perhaps explain the claimed paucity of short-period planets around rapidly rotating stars \citep{McQuillanMA13, WalkowiczB13}.
Certainly current observations (e.g., Table ~1 in VR14b and references therein) suggest that the host stars of hot Jupiters close to $a_{\rm R}$ are all slow rotators.

As a natural continuation of our previous work (VR14, VR14b), we have used a standard binary MT model to 
investigate the evolution of a hot Jupiter undergoing RLO. In particular we focus here on the evolution of the planetary mass and 
orbital separation during stable MT.
Indeed, for typical systems where a hot Jupiter orbits a solar-like star, we expect the MT to be dynamically stable. Based on \cite{SepinskyWKR10} the initial mass stream leaving the planet near the inner Lagrange point L1 will not
impact the surface of the star, so MT will proceed through an accretion disk. This is the standard case for close binary stars with a very small donor-to-accretor mass ratio, and it is indeed expected to be dynamically 
stable \citep{Metzger+12}. This is also in agreement with \citeauthor{LaiHvdH2010}'s (\citeyear{LaiHvdH2010}) detailed study for WASP-12.

As in our previous work we use full stellar evolution models and a detailed treatment of tidal dissipation.
In what follows $M_{\rm pl}$, $R_{\rm pl}$, and $M_{\rm c}$ are the planetary mass, radius, and core mass, respectively. The stellar mass, radius, spin (orbital) frequency, and main-sequence lifetime are $M_{\rm *}$, $R_{*}$, $\Omega_{*}$ ($\Omega_{o}$), and $t_{\rm MS}$, respectively, while the orbital period is $P_{\rm orb}$, and the mass ratio is $q=M_{*}/M_{\rm pl}$. 

\section{Observational Motivation}\label{Observations}

This study is partly motivated by the excess of \kepler\ single-candidate systems with sizes less than a few Earth radii ($R_{\oplus}$) and $P_{\rm orb}$ less than a few days that was seen in \citet{Steffen:2013c}. They noted a significant excess of isolated hot super-Earth- or sub-Neptune-size planets, stating that they might be a small-planet analog of the hot Jupiters. 

\begin{figure} [!h]
\epsscale{1.2}
\plotone{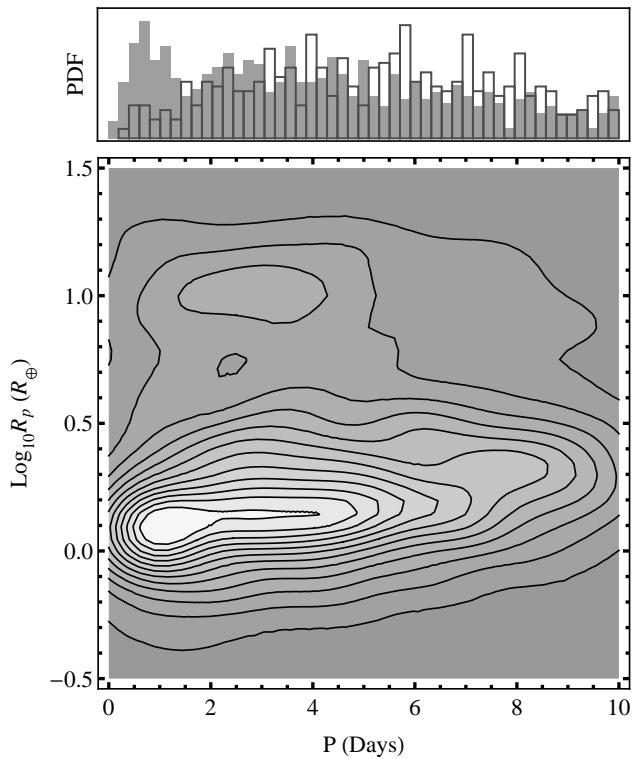}
\caption{Upper panel: Probability densities of orbital period for single KOIs (grey) and multiple KOIs (outline) for planets with sizes less than 5 \re .  These distributions are unlikely to be from the same parent distribution, primarily due to the excess of single-planet systems with orbits near 1 day.  Lower panel: The distribution of planet sizes and orbital periods for single KOI systems.  The hot Jupiter and hot super-Earth populations are visible as peaks near 10 \re\ and 3-days and 1 \re\ and 1-day respectively.}
\label{fig:singlevsmultis} 
\end{figure}

Figure~\ref{fig:singlevsmultis} shows a histogram comparing the orbital period distribution of \kepler\ objects of interest (KOIs) with sizes $<$5\,\re\,\,and $P_{\rm orb}<$10 days for single- and multi-planet systems using data from the Quarter 8 (Q8) catalog \citep{Burke:2014}.  Both a Kolmogorv-Smirnov test and an Anderson-Darling test yield $p$-values  $\sim10^{-7}$ indicating a significant difference between the period distributions of single- and multi-planet systems in this regime.  False positive signals are unlikely to have any impact on the statistical significance of this excess (see \citet{Steffen:2013c} for a discussion).  

Also shown in Figure~\ref{fig:singlevsmultis} is the distribution of $P_{\rm orb}$ and $R_{\rm pl}$ for \kepler\ single-planet candidate systems (generated through standard Gaussian smoothing using Silverman's rule---default options in Mathematica).  One can identify the island of hot Jupiters centered near 3 days and 10 \re\, as well as the peak near 1 day and 1 \re.
This excess population of low-mass single planets is the one we are trying to explain here, as these objects could be the remnants of hot Jupiters
that have lost their envelopes through RLO.

\section{Orbital Evolution Model}\label{Orbital Evolution Model}
Our assumptions are based on the properties of observed hot Jupiters close to $a_{\rm R}$ (VR14b).  We consider circular orbits and assume that the star is slowly rotating ($\Omega_{*}\ll \Omega_{o}$) and the stellar spin aligned with the orbital angular momentum---consistent with the majority of systems with solar-type stars. We assume the planet to be tidally locked.
Our models (VR14, VR14b) can account for stellar wind mass loss and magnetic braking, but these mechanisms do not impact our results 
significantly and we omit them for simplicity.
Another mechanism that might affect the evolution of the tightest hot Jupiters is photo-evaporation (e.g., the bright HD 209458b might be shedding mass through this mechanism, \citealt{VidalMadjar2004}). \cite{Murray-Clay+09} quoted a maximum mass-loss rate of  $3.3\times\,10^{10}\,{\rm g}\,{\rm s}^{-1}$ and noted that a hot Jupiter cannot lose a significant fraction of its mass via photo-evaporation at any stage during its lifetime. 
We included this upper limit in our models before the envelope is completely lost and found that indeed most of the orbital evolution is not significantly affected. Therefore, we do not consider photo-evaporation further here (but see Section~\ref{Conclusions}).

We account for the host star evolution using MESA (version  4798, \citealt{PBDHLT2011, Paxton+13}) as in VR14. 
Our variables are $a$, $\Omega_{*}$, $M_{\rm pl}$ (donor) and $M_{*}$ (accretor), and their evolution is described by
\begin{align}
&\dot{a} = \dot{a}_{\rm tid}+\dot{a}_{\rm MT} \label{eq:aDotTot};\\
&\dot{\Omega}_{\rm *} = \dot{\Omega}_{\rm *, tid}+\dot{\Omega}_{*, \rm evol};\label{eq:OmegaDotTot}\\
&\dot{M}_{\rm pl} = \dot{M}_{\rm pl, MT};\label{eq:M2DotTot}\\
&\dot{M}_{\rm *} = \dot{M}_{\rm *, MT}\label{eq:M1DotTot};.
\end{align}

The subscripts ``tid'', ``MT'', and ``evol'' refer to  tides, MT, and stellar evolution, respectively. The terms entering Equation~(\ref{eq:aDotTot}) can 
be derived from the system's total angular momentum, 
\begin{align}
L = G^{1/2}(M_{\rm *}+M_{\rm pl})^{-1/2}M_{\rm *}M_{\rm pl}a^{1/2}, 
\label{eq:L}
\end{align}
where $G$ is the gravitational constant, as follows.
Let $\beta$ represent the fraction of mass lost by the planet and accreted onto the star via MT, then
\begin{align}
\dot{M}_{\rm *, MT} = -\beta\dot{M}_{\rm pl, MT}.
\label{eq:M1DotMT}
\end{align}

The total change in $L$ is given by
\begin{align}
&\frac{\dot{L}}{L} =\left[\frac{\beta-1}{2(M_{\rm *}+M_{\rm pl})}+\left(\frac{1}{M_{\rm pl}}-\frac{\beta}{M_{\rm *}}\right)\right]\dot{M}_{\rm pl,MT}+
\frac{(\dot{a}_{\rm MT}+\dot{a}_{\rm tid})}{2a},
\label{eq:JorbTotalDeriv1}
\end{align}
or
\begin{align}
\frac{\dot{L}}{L} =\frac{1}{L}(\dot{L}_{\rm MT}+\dot{L}_{\rm tid}), 
\label{eq:JorbTotalDerivWhereItIsLostFrom}
\end{align}
where $\dot{L}_{\rm MT}$ and $\dot{L}_{\rm tid}$ represent the change in $L$ with respect to the system center of mass.
Following the standard binary star treatment (e.g., \citealt{Rappaport+82}), 
we assume that the angular momentum carried onto the accretion disk is returned to the orbit via tidal torques (e.g., \citealt{PriedhorskyV88}).  Thus, $\dot{a}_{\rm MT}$ can be derived introducing the angular momentum parameter $\alpha$ defined as
\begin{align}
\dot{L}_{\rm MT}= \alpha\dot{M}_{\rm pl,MT}(1-\beta)L\frac{(M_{\rm *}+M_{\rm pl})}{M_{*}M_{\rm pl}} \label{eq:LorbMT}.
\end{align}
From Equations~(\ref{eq:JorbTotalDeriv1}) and (\ref{eq:JorbTotalDerivWhereItIsLostFrom}) it follows that
\begin{align}
\frac{\dot{a}_{\rm MT}}{a}&=-2\frac{\dot{M}_{\rm pl, MT}}{M_{\rm pl}}\left(1-\frac{1}{q}\right)\label{eq:aDotMT}, 
\end{align}
 where we have set $\beta\,=\,1$ (conservative MT, where $\alpha$ is irrelevant) for simplicity. In the limit of large $q$ the final $a-M_{\rm pl}$ relation is independent on $\alpha$ and $\beta$ (i.e., the details of the MT process), as they only affect the {\it duration} of the RLO phase.
For $\dot{a}_{\rm tid}$, we treat stellar tides in the weak-friction approximation \citep{Zahn1977, Zahn1989}, as in VR14 and VR14b. Specifically, we assume that the tidal perturbation is dissipated via eddy viscosity operating in the stellar convection zone (Equations~(1), (4), (6), and (7) in VR14). For the reduction in the efficiency of tides at high tidal forcing frequencies, we use the linear theory of \cite{Zahn1966}, which is consistent with recent numerical results \citep{PenevSRD2007}. As we consider only sub-synchronous configurations ($\Omega_{*}/\Omega_{\rm o}<1$), tidal dissipation leads to orbital decay.

The planetary mass evolution due to RLO is derived as in \cite{Rappaport+82}, where this phase begins when the planet fills its Roche lobe  (of radius $R_{\rm L}$), and continues as long as $\dot{R}_{\rm pl}\,=\,\dot{R}_{\rm L}$. For large $q$ \citep{Paczynski71} 
\begin{equation}
R_{\rm L} =  a\left(\frac{2}{3^{4/3}}\right)(1+q)^{-1/3}.
\label{eq:RLO}
\end{equation}
To determine $\dot{M}_{\rm pl,MT}$, a mass--radius relation is needed. Detailed models suggest that the thermal timescale of typical hot Jupiters is $<\,$1\,Myr (e.g.,  Fig.~2 in \citealt{SpiegelBurrows12}), while the RLO timescales $M_{\rm pl}/\dot{M}_{\rm pl,MT}$ computed here are almost always $\gg 1\,$Myr. 
Therefore, we assume that throughout RLO the planet remains in thermal equilibrium (but see Section~\ref{Conclusions}). 
We have used two different approximations for the planet: a simple $n\,=\,1$ polytrope, corresponding to a constant radius independent of mass, and more
realistic models with rocky cores, fitted to the results of \citet{BatyginStevenson13} and \cite{FortneyMB07}. Specifically, we consider models 
with core masses $M_{\rm c}\,=\,1M_{\oplus}, 3M_{\oplus}$, and~$10M_{\oplus}$. 

\begin{figure} [!h]
\epsscale{1.2}
\plotone{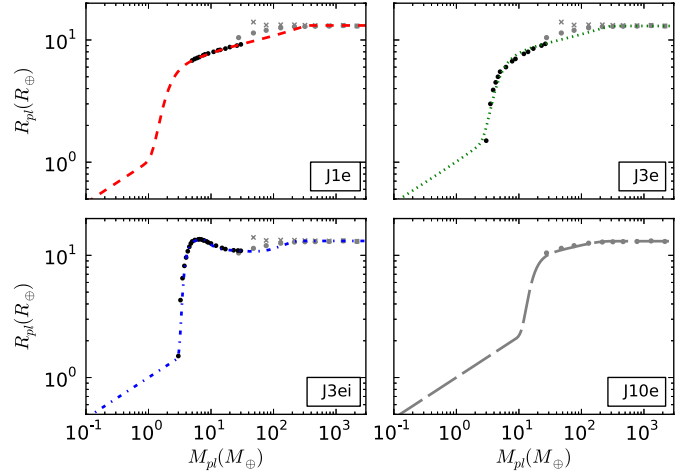}
\caption{Mass--radius relations for the planet. In black are data taken from panels A and B of Fig.~3 in \cite{BatyginStevenson13}, while in grey are the data taken from Table~4 of \cite{FortneyMB07} for a giant planet with no core (crosses) and  $M_{\rm c}=10\,M_{\oplus}$ (filled circles). The labels ``J1e'', ``J3e'', and ``J10e''
refer to the 1$\,M_{\oplus}$, 3$\,M_{\oplus}$, and 10$\,M_{\oplus}$ core-mass models, while the label ``J3ei'' denotes the irradiated 3$\,M_{\oplus}$ core-mass model.}
\label{fig:massAndRadius_Batygin}
\end{figure}

For the low envelope masses reached near the end of the RLO phase we also consider qualitatively the effects of strong irradiation,
as modeled in detail by \citet{BatyginStevenson13}.
Note that the simple $n=1$ polytrope is actually a reasonably good approximation for large envelope masses, as detailed models
show a nearly constant radius as a function of mass (e.g., Fig.~3 in \citealt{BatyginStevenson13} or Fig.~8 in \citealt{FortneyMB07} for the high-mass end). 
Below we denote the 1$\,M_{\oplus}$, 3$\,M_{\oplus}$, and 10$\,M_{\oplus}$ core-mass models as ``J1e,'' ``J3e,'' and ``J10e,'' respectively. We denote the irradiated 3$\,M_{\oplus}$ core-mass model as ``J3ei.'' These are shown in Fig.~\ref{fig:massAndRadius_Batygin} and are described by the following equations in Earth units.
For J1e and J3e, 
\begin{small}
\begin{align}
\label{eq:Batygin}
R_{\rm pl}&=A{\rm e}^{-\frac{B}{M_{\rm pl}^{C}}} + M_{\rm pl}^{\frac{1}{3}}(1 - M_{\rm pl}^{0.01}{\rm e}^{-\frac{D}{M_{\rm pl}^{C}}})+\frac{M_{\rm pl}^{2}}{100+M_{\rm pl}^{2}},
\end{align}
\end{small}
where for J1e (J3e) A = 5.2 (5.5), B = 5 (70), C = 2.7 (3), and D = $3\times\,10^{7}$ ($1\times\,10^{8}$). 
For J3ei, 
\begin{small}
\begin{align}
\label{eq:BatyginIrr}
R_{\rm pl}&=11.6\,{\rm e}^{-\frac{1\times 10^{5}}{ M_{\rm pl}^{9}}} + M_{\rm pl}^{\frac{1}{3}}(1 - M_{\rm pl}^{0.07} {\rm e}^{-\frac{5\times 10^{5}}{M_{\rm pl}^{6}}} )+ \frac{10M_{\rm pl}^{2.5}}{5\times 10^{5}+M_{\rm pl}^{2.6}}, 
\end{align}
\end{small}
while for J10e  we take
\begin{small}
\begin{align}
\label{eq:Fortney}
R_{\rm pl}&=7.5\,{\rm e}^{-\frac{5\times 10^{4}}{M_{\rm pl}^{4}}}+ M_{\rm pl}^{\frac{1}{3}}(1 -  {\rm e}^{-\frac{3\times 10^{2}}{M_{\rm pl}^{0.9}}})
\end{align}
\end{small}
Here we have assumed a constant density profile ($R_{\rm pl}\propto\,M_{\rm pl}^{1/3}$) below the core mass. The above expression all have a maximum at $M_{\rm pl, max}\simeq\,1M_{\rm J}$. We use the fits in Equations~(\ref{eq:Batygin})--(\ref{eq:Fortney}) for $M_{\rm pl}<M_{\rm pl, max}$ and consider a constant radius profile for  $M_{\rm pl}\geq\,M_{\rm pl, max}$.

Setting $\dot{R}_{\rm pl}\,=\,\dot{R}_{\rm L}$ yields
\begin{equation}
\frac{\dot{M}_{\rm pl,MT}}{M_{\rm pl}} = \frac{\frac{\dot{a}_{\rm tid}}{2a}}{\frac{5}{6}-\frac{1}{q}+\frac{\xi}{2}}
\label{eq:dotM2MT}
\end{equation}
for a polytrope of index $n$, where $\xi\,=\frac{d{\rm ln}R_{\rm PL}}{d{\rm ln}M_{\rm PL}}=\,\frac{n-1}{n-3}$. A more complicated expression exists for our detailed models and it can be generalized to any mass--radius relation. Hereafter, when mentioning the polytrope, we assume $n=1$ ($\xi\,=\,0$).
\cite{Rappaport+82} note that a necessary condition for stable MT is that the denominator in Equation~(\ref{eq:dotM2MT}) be positive.
For non-conservative MT the condition for stability reduces to $\alpha(1-\beta)<\frac{5}{6}+\frac{\xi}{2}$ when $q >>1$. If mass is lost from the system with the specific angular momentum of the donor ($\alpha\,=\,1$), the RLO phase is stable if $\beta\,\gtrsim\,0.17$ for the polytrope. 
The parameter space for stability widens for more realistic planetary models with core, provided that $\xi>0$ (neglecting irradiation).
The case $\alpha=1$ and $\beta<1$ 
might occur because of tidally enhanced mass loss through photo-evaporation \citep{Jackson+10,LeitzingerOKLW11} or magnetically controlled outflows from the planet \citep{Adams2011,CohenGlocer12}.

In addition, we check for stellar RLO \citep{Eggleton1983} and stop our calculation whenever it occurs, as it will lead to rapid engulfment of the planet's remnant 
by the star.

A trivial relation for $a$ as a function of $M_{\rm pl}$ is obtained in the limit $q>>1$. Combining Equations~(\ref{eq:aDotTot}), (\ref{eq:aDotMT}), and (\ref{eq:dotM2MT}), and considering a polytrope, we find
\begin{equation}
\frac{a}{a_{0}} = \left(\frac{M_{\rm pl}}{M_{\rm pl, 0}}\right)^{-1/3}.
\label{eq:aAnalyticMass}
\end{equation}
where the subscript ``0'' denotes the values at the onset of RLO. An analogous expression can be derived for our detailed models. 


\section{Examples}\label{Example}
We now illustrate how hot Jupiters can naturally evolve into lower-mass planets. 
We take a 
2\,$M_{\rm J}$ hot Jupiter (so that $M_{\rm pl}$ is in the plateau of Fig.~\ref{fig:massAndRadius_Batygin}). For all models we compute $R_{\rm pl}$ from the maxima of Equations~(\ref{eq:Batygin})--(\ref{eq:Fortney}),  giving $\simeq$1.2\,$R_{\rm J}$. The host star has a mass of 1$\,M_{\odot}$ and solar metallicity. 
Our initial conditions are $a\,=\,1.5\,a_{\rm R}\,\simeq\,0.016\,$AU ($P_{\rm orb}\,\simeq\,$0.65\,days) and $\Omega_{*}\,=\,0.1\,\Omega_{\rm o}$ (our results change little varying $\Omega_{*}/\Omega_{\rm o}$ between 0--0.15, consistent with observations). We start the orbital evolution (arbitrarily)
when the stellar age is 0.3\,$t_{\rm MS}$ ($\sim\,$3\,Gyr) and consider both conservative and non-conservative MT, with $\alpha=1$ for the latter. For the non-conservative MT case, we illustrate extreme examples with $\beta\,=\,$0.2 for the polytrope, J1e, J3e, and J10e, and $\beta\,=\,$0.4 for J3ei. These values are at the limit for MT stability (Equation~(\ref{eq:dotM2MT})).
\begin{figure} [!h]
\epsscale{1.2}
\plotone{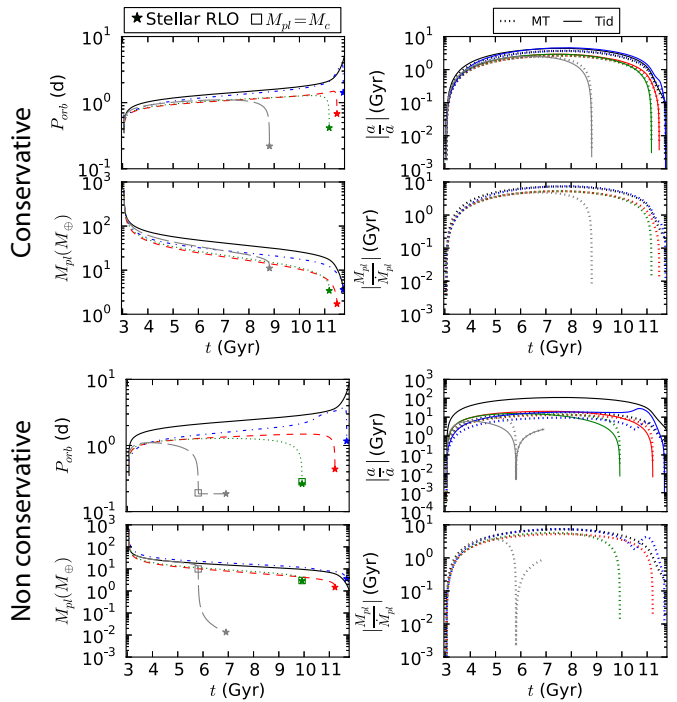}
\caption{Orbital evolution examples for conservative (top four panels) and non-conservative (bottom four panels) MT. For the latter $\beta\,=\,$0.2 for the polytrope, J1e, J3e, and J10e, and $\beta\,=\,$0.4 for J3ei, while $\alpha=1$. The different models are identified by color and line style on the left (as in Fig.~\ref{fig:massAndRadius_Batygin}) and color on the right. Black lines are for the polytrope. For each set of four panels on the left are $P_{\rm orb}$ (top) and $M_{\rm pl}$ (bottom), while on the right are the timescales for the evolution of $a$ (top) and $M_{\rm pl}$ (bottom) due to MT (dotted lines) and tides (solid lines). }
\label{fig:evolutionExample}
\end{figure} 
As shown in Fig.~\ref{fig:evolutionExample}, prior to MT the orbit decays fast as tides remove orbital angular momentum to spin up the star. After only $\sim$4\,Myr RLO begins and the orbit now expands. For the polytrope and J3ei models, the flat mass--radius relation (Fig.~\ref{fig:massAndRadius_Batygin}) leads to significant orbital expansion. For J3ei the orbit expands as long as MT dominates over tides. Eventually, as the star approaches the end of the main sequence, the increase in $R_{*}$ and the dependence of $\dot{a}_{\rm tid}$ on $(R_{*}/a)^{8}$ (Equations~(1) and (4) in VR14) cause tides to take over the MT and the system to evolve more rapidly. The evolution stops when stellar RLO begins (denoted with ``$\bigstar$''). The evolution differs for J1e, J3e, and J10e. In fact, while MT always causes some orbital expansion at the onset of RLO, tides take over sooner when the mass--radius relation steepens. The orbit begins shrinking, consuming the planet faster. If the MT is conservative, the evolution stops because of stellar RLO. The same is true for non-conservative MT but the cores of J3e and J10e are exposed prior to stellar RLO (marked by ``$\square$''). For J10e in particular, once the core is exposed $a$ remains constant, as expected for a constant density model ($R_{\rm pl}\propto\,M_{\rm pl}^{1/3}$). 
Even without magnetic braking, at the end of the calculation the star is spinning at less than 10\% break-up in all cases.
\section{Comparison With Observations}\label{Comparison With Observations}
\begin{figure} [!h]
\epsscale{1.1}
\plotone{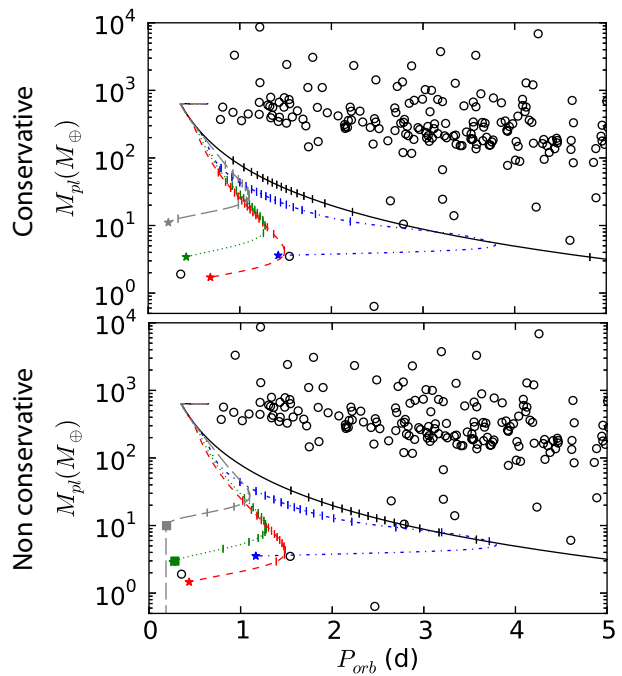}
\caption{Confirmed single planets with known $M_{\rm pl}$, $R_{\rm pl}$, and $P_{\rm orb}$ (black open circles) within the range displayed. 
The colors and line styles are as in Fig.~\ref{fig:massAndRadius_Batygin}. The vertical tick marks on each line denote intervals of 500\,Myr.}
\label{fig:Radius_Porb_KOI_confirmed} 
\end{figure}
Fig.~\ref{fig:Radius_Porb_KOI_confirmed} shows the mass and orbital period of the known single planets with observationally inferred $M_{\rm pl}$, $R_{\rm pl}$, and $P_{\rm orb}$ (NASA Exoplanet Archive, 25 April 2014), and the evolutionary tracks of Fig.~\ref{fig:evolutionExample}. While most of the mass is lost within a few Gyr, the orbital evolution slows down as the orbit expands (see also Fig.~\ref{fig:evolutionExample}, right), and it eventually accelerates when tides dominate over MT.

Varying the initial $M_{\rm pl}$ while using the same initial conditions and planetary models, the orbital evolution begins and ends at different $P_{\rm orb}$, but it continues along the same tracks displayed in Fig.~\ref{fig:Radius_Porb_KOI_confirmed} (e.g., starting with a higher $M_{\rm pl}$, once $M_{\rm pl}$ drops to 2$M_{\rm J}$, the evolutionary tracks overlap). Furthermore, varying $M_{*}$ between 0.5$M_{\odot}$-2$M_{\odot}$ (typical for \kepler\,\,targets) changes the duration of the various phases displayed in Fig.~\ref{fig:Radius_Porb_KOI_confirmed}, but it does not affect the shape of the evolutionary tracks significantly. 
In our examples the evolution prior to RLO lasts only $\sim\,$ 4Myr, which may seem at odds with the observed hot Jupiters close to $a_{\rm R}$. However, the evolutionary timescales can easily vary over several orders of magnitude depending both on the star under consideration as well as the assumed tidal efficiency. For instance, the pre-RLO phase for a host star of 0.5$M_{\odot}$ lasts several tens of Myrs because of the smaller $R_{*}$. Furthermore, debate still exists on whether the reduction in the effectiveness of tides at high tidal forcing frequencies should be described by a linear  (adopted here) or quadratic prescription \citep{GoldreichNicholson1977}. The latter yields longer orbital evolution timescales, with a duration of the pre-RLO phase of $\sim$\,100\,Myr ($\sim$\,10\,Gyr) for a 1$M_{\odot}$ (0.5 $M_{\odot}$) host star. 

Interestingly, some of the few known Earth-size planets with observationally inferred masses (e.g., the intriguing Kepler-78, with $P_{\rm orb}\simeq\,0.35\,$days, or Kepler-98 at $P_{\rm orb}\simeq\,1.5\,$days and $M_{\rm pl}\simeq\,3.5M_{\oplus}$) lie along or very close to our evolutionary tracks. Clearly, more mass measurements in this short orbital period and small mass regime would provide an important test of these ideas.
\section{Discussion} \label{Conclusions}
Our simple models naturally leads to planets similar in orbital period and mass to {\it known\/} systems hosting very hot sub-Neptunes and super-Earths.
Additionally, the wide range of masses and orbital periods covered by our evolutionary tracks suggest that, considering a variety of planetary interior models, the excess short-period, Earth-sized planets seen in the \kepler\ data \citep{Steffen:2013c} could be explained as being the remnant planet cores from hot Jupiters that went through RLO and lost their gaseous envelopes. 
One implication is that the number of systems that have ever had a hot Jupiter is likely to be about 3 times larger than what one would infer directly from observations, since the number of single hot Earths and super-Earths in the \kepler\ data is nearly twice the number of hot Jupiters.
If correct, these results suggest that many Jovian planets have rocky cores and give a means to study the properties of those cores directly.  

Core-less Jupiters (modeled here as simple $n=1$ polytropes) would appear as very low-density planets at orbital periods of a few days and with sizes near their Roche lobe.  An absence of such planets would imply that most or all gas giants that form in systems capable of producing hot Jupiters have rocky cores.
Observationally, if the excess hot Earths comes from hot Jupiters, then the host stars  of these populations should have similar properties.  Specifically, their metallicities and masses should be somewhat higher than solar.  

Depending on the details of the planetary interior, the orbital evolution timescales could be long enough that it might be possible to observe planets currently undergoing RLO.
As shown by \cite{LaiHvdH2010} the resulting accretion disk could 
produce line absorption of stellar radiation, time-dependent obscuration of the starlight, and an earlier ingress for planetary transits.

Here we adopted simplified models for the planets, assuming thermal equilibrium throughout and using simple fits to published mass-radius relations. We 
also neglected photo-evaporation, even though it could play an important role in the evolution of highly irradiated super-Earths and sub-Neptunes (e.g., \citealt{BatyginStevenson13}; \citealt{LopezFortney13} and references therein). For example, it could naturally explain the density contrast observed in Kepler-36 \citep{LopezFortney13}.
If, near the end of the evolution, mass loss from the planet were significantly enhanced through photo-evaporation,
this could lead to orbital expansion, potentially slowing down MT, or even completely halting RLO.
Magnetic fields and stellar winds, also ignored here, may well play a role in both enhancing photo-evaporation and in channeling or entraining the MT flow
\citep{CohenGlocer12}, questioning whether the MT really proceeds through an accretion disk, with all the angular momentum being returned to the orbit.
Finally, even though we assumed that the planet's spin remains tidally locked throughout the evolution, the synchronous rotation cannot strictly be maintained as the orbit changes significantly. This could lead to significant tidal heating, potentially advancing the onset of RLO  \citep{Hansen2012} and enhancing mass loss. 
These various effects will have to be studied carefully in future work, which could combine the simple treatment of orbital evolution and tides introduced here
with more detailed models of the planet (e.g., using MESA, as in \citealt{BatyginStevenson13}).
This approach, to use detailed models for the donor in MT calculations which are computed self-consistently as RLO proceeds, 
has been applied successfully for many years in calculations of interacting binary stars (e.g., \citealt{Madhusudhan+08}).

 \acknowledgments

\begin{acknowledgements}
FV and FAR are supported by NASA Grant NNX12AI86G. JHS is supported by NASA Grant NNH12ZDA001N-KPS. We thank Nick Cowan, Vicky Kalogera, Saul Rappaport, Brian Metzger, and Brad Hansen for useful discussions.
This research  made use of the NASA Exoplanet Archive.
\end{acknowledgements}


\begin{thebibliography}{}
\expandafter\ifx\csname natexlab\endcsname\relax\def\natexlab#1{#1}\fi

\bibitem[{{Adams}(2011)}]{Adams2011}
{Adams}, F.~C. 2011, \apj, 730, 27

\bibitem[{{Albrecht} {et~al.}(2012){Albrecht}, {Winn}, {Johnson}, {Howard},
  {Marcy}, {Butler}, {Arriagada}, {Crane}, {Shectman}, {Thompson}, {Hirano},
  {Bakos}, \& {Hartman}}]{Albrecht+12}
{Albrecht}, S., {Winn}, J.~N., {Johnson}, J.~A., {et~al.} 2012, \apj, 757, 18

\bibitem[{{Batygin} \& {Stevenson}(2013)}]{BatyginStevenson13}
{Batygin}, K., \& {Stevenson}, D.~J. 2013, \apjl, 769, L9

\bibitem[{{Burke} {et~al.}(2014){Burke}, {Bryson}, {Mullally}, {Rowe},
  {Christiansen}, {Thompson}, {Coughlin}, {Haas}, {Batalha}, {Caldwell},
  {Jenkins}, {Still}, {Barclay}, {Borucki}, {Chaplin}, {Ciardi}, {Clarke},
  {Cochran}, {Demory}, {Esquerdo}, {Gautier}, {Gilliland}, {Girouard}, {Havel},
  {Henze}, {Howell}, {Huber}, {Latham}, {Li}, {Morehead}, {Morton}, {Pepper},
  {Quintana}, {Ragozzine}, {Seader}, {Shah}, {Shporer}, {Tenenbaum}, {Twicken},
  \& {Wolfgang}}]{Burke:2014}
{Burke}, C.~J., {Bryson}, S.~T., {Mullally}, F., {et~al.} 2014, \apjs, 210, 19

\bibitem[{{Chang} {et~al.}(2010){Chang}, {Gu}, \&
  {Bodenheimer}}]{ChangGuBodenheimer10}
{Chang}, S.-H., {Gu}, P.-G., \& {Bodenheimer}, P.~H. 2010, \apj, 708, 1692

\bibitem[{{Cohen} \& {Glocer}(2012)}]{CohenGlocer12}
{Cohen}, O., \& {Glocer}, A. 2012, \apjl, 753, L4

\bibitem[{{Eggleton}(1983)}]{Eggleton1983}
{Eggleton}, P.~P. 1983, \apj, 268, 368

\bibitem[{{Fabrycky} \& {Tremaine}(2007)}]{FabryckyTremaine07}
{Fabrycky}, D., \& {Tremaine}, S. 2007, \apj, 669, 1298

\bibitem[{{Ford} \& {Rasio}(2006)}]{FordRasio06}
{Ford}, E.~B., \& {Rasio}, F.~A. 2006, \apjl, 638, L45

\bibitem[{{Fortney} {et~al.}(2007){Fortney}, {Marley}, \&
  {Barnes}}]{FortneyMB07}
{Fortney}, J.~J., {Marley}, M.~S., \& {Barnes}, J.~W. 2007, \apj, 659, 1661

\bibitem[{{Goldreich} \& {Nicholson}(1977)}]{GoldreichNicholson1977}
{Goldreich}, P., \& {Nicholson}, P.~D. 1977, Icarus, 30, 301

\bibitem[{{Hansen}(2012)}]{Hansen2012}
{Hansen}, B.~M.~S. 2012, \apj, 757, 6

\bibitem[{{Jackson} {et~al.}(2009){Jackson}, {Barnes}, \&
  {Greenberg}}]{Jackson+09}
{Jackson}, B., {Barnes}, R., \& {Greenberg}, R. 2009, \apj, 698, 1357

\bibitem[{{Jackson} {et~al.}(2008){Jackson}, {Greenberg}, \&
  {Barnes}}]{Jackson+08}
{Jackson}, B., {Greenberg}, R., \& {Barnes}, R. 2008, \apj, 678, 1396

\bibitem[{{Jackson} {et~al.}(2010){Jackson}, {Miller}, {Barnes}, {Raymond},
  {Fortney}, \& {Greenberg}}]{Jackson+10}
{Jackson}, B., {Miller}, N., {Barnes}, R., {et~al.} 2010, \mnras, 407, 910

\bibitem[{{Lai}(2012)}]{Lai12}
{Lai}, D. 2012, \mnras, 423, 486

\bibitem[{{Lai} {et~al.}(2010){Lai}, {Helling}, \& {van den
  Heuvel}}]{LaiHvdH2010}
{Lai}, D., {Helling}, C., \& {van den Heuvel}, E.~P.~J. 2010, \apj, 721, 923

\bibitem[{{Leitzinger} {et~al.}(2011){Leitzinger}, {Odert}, {Kulikov},
  {Lammer}, {Wuchterl}, {Penz}, {Guarcello}, {Micela}, {Khodachenko},
  {Weingrill}, {Hanslmeier}, {Biernat}, \& {Schneider}}]{LeitzingerOKLW11}
{Leitzinger}, M., {Odert}, P., {Kulikov}, Y.~N., {et~al.} 2011, \planss, 59,
  1472

\bibitem[{{Lopez} \& {Fortney}(2013)}]{LopezFortney13}
{Lopez}, E.~D., \& {Fortney}, J.~J. 2013, \apj, 776, 2

\bibitem[{{Madhusudhan} {et~al.}(2008){Madhusudhan}, {Rappaport},
  {Podsiadlowski}, \& {Nelson}}]{Madhusudhan+08}
{Madhusudhan}, N., {Rappaport}, S., {Podsiadlowski}, P., \& {Nelson}, L. 2008,
  \apj, 688, 1235

\bibitem[{{Matsumura} {et~al.}(2010){Matsumura}, {Peale}, \&
  {Rasio}}]{MatsumuraPR2010}
{Matsumura}, S., {Peale}, S.~J., \& {Rasio}, F.~A. 2010, \apj, 725, 1995

\bibitem[{{McQuillan} {et~al.}(2013){McQuillan}, {Mazeh}, \&
  {Aigrain}}]{McQuillanMA13}
{McQuillan}, A., {Mazeh}, T., \& {Aigrain}, S. 2013, \apjl, 775, L11

\bibitem[{{Metzger} {et~al.}(2012){Metzger}, {Giannios}, \&
  {Spiegel}}]{Metzger+12}
{Metzger}, B.~D., {Giannios}, D., \& {Spiegel}, D.~S. 2012, \mnras, 425, 2778

\bibitem[{{Murray-Clay} {et~al.}(2009){Murray-Clay}, {Chiang}, \&
  {Murray}}]{Murray-Clay+09}
{Murray-Clay}, R.~A., {Chiang}, E.~I., \& {Murray}, N. 2009, \apj, 693, 23

\bibitem[{{Nagasawa} {et~al.}(2008){Nagasawa}, {Ida}, \& {Bessho}}]{Nagasawa08}
{Nagasawa}, M., {Ida}, S., \& {Bessho}, T. 2008, \apj, 678, 498

\bibitem[{{Naoz} {et~al.}(2011){Naoz}, {Farr}, {Lithwick}, {Rasio}, \&
  {Teyssandier}}]{Naoz+11}
{Naoz}, S., {Farr}, W.~M., {Lithwick}, Y., {Rasio}, F.~A., \& {Teyssandier}, J.
  2011, \nat, 473, 187

\bibitem[{{Paczy{\'n}ski}(1971)}]{Paczynski71}
{Paczy{\'n}ski}, B. 1971, \araa, 9, 183

\bibitem[{{Paxton} {et~al.}(2011){Paxton}, {Bildsten}, {Dotter}, {Herwig},
  {Lesaffre}, \& {Timmes}}]{PBDHLT2011}
{Paxton}, B., {Bildsten}, L., {Dotter}, A., {et~al.} 2011, \apjs, 192, 3

\bibitem[{{Paxton} {et~al.}(2013){Paxton}, {Cantiello}, {Arras}, {Bildsten},
  {Brown}, {Dotter}, {Mankovich}, {Montgomery}, {Stello}, {Timmes}, \&
  {Townsend}}]{Paxton+13}
{Paxton}, B., {Cantiello}, M., {Arras}, P., {et~al.} 2013, \apjs, 208, 4

\bibitem[{{Penev} {et~al.}(2007){Penev}, {Sasselov}, {Robinson}, \&
  {Demarque}}]{PenevSRD2007}
{Penev}, K., {Sasselov}, D., {Robinson}, F., \& {Demarque}, P. 2007, \apj, 655,
  1166

\bibitem[{{Plavchan} \& {Bilinski}(2013)}]{PlavchanBilinski13}
{Plavchan}, P., \& {Bilinski}, C. 2013, \apj, 769, 86

\bibitem[{{Priedhorsky} \& {Verbunt}(1988)}]{PriedhorskyV88}
{Priedhorsky}, W.~C., \& {Verbunt}, F. 1988, \apj, 333, 895

\bibitem[{{Rappaport} {et~al.}(1982){Rappaport}, {Joss}, \&
  {Webbink}}]{Rappaport+82}
{Rappaport}, S., {Joss}, P.~C., \& {Webbink}, R.~F. 1982, \apj, 254, 616

\bibitem[{{Rasio} \& {Ford}(1996)}]{RasioFord96}
{Rasio}, F.~A., \& {Ford}, E.~B. 1996, Science, 274, 954

\bibitem[{{Schlaufman} \& {Winn}(2013)}]{SchlaufmanWinn13}
{Schlaufman}, K.~C., \& {Winn}, J.~N. 2013, \apj, 772, 143

\bibitem[{{Sepinsky} {et~al.}(2010){Sepinsky}, {Willems}, {Kalogera}, \&
  {Rasio}}]{SepinskyWKR10}
{Sepinsky}, J.~F., {Willems}, B., {Kalogera}, V., \& {Rasio}, F.~A. 2010, \apj,
  724, 546

\bibitem[{{Spiegel} \& {Burrows}(2012)}]{SpiegelBurrows12}
{Spiegel}, D.~S., \& {Burrows}, A. 2012, \apj, 745, 174

\bibitem[{{Steffen} \& {Farr}(2013)}]{Steffen:2013c}
{Steffen}, J.~H., \& {Farr}, W.~M. 2013, \apjl, 774, L12

\bibitem[{{Teitler} \& {K{\"o}nigl}(2014)}]{TeitlerKonigl14}
{Teitler}, S., \& {K{\"o}nigl}, A. 2014, ArXiv e-prints, arXiv:1403.5860

\bibitem[{{Trilling} {et~al.}(1998){Trilling}, {Benz}, {Guillot}, {Lunine},
  {Hubbard}, \& {Burrows}}]{TrillingBGLHB1998}
{Trilling}, D.~E., {Benz}, W., {Guillot}, T., {et~al.} 1998, \apj, 500, 428

\bibitem[{{Valsecchi} \& {Rasio}(2014{\natexlab{a}})}]{ValsecchiR+14}
{Valsecchi}, F., \& {Rasio}, F.~A. 2014{\natexlab{a}}, \apj, 786, 102

\bibitem[{{Valsecchi} \& {Rasio}(2014{\natexlab{b}})}]{Valsecchi+14edge}
---. 2014{\natexlab{b}}, \apjl, 787, L9

\bibitem[{{Vidal-Madjar} {et~al.}(2004){Vidal-Madjar}, {D{\'e}sert},
  {Lecavelier des Etangs}, {H{\'e}brard}, {Ballester}, {Ehrenreich}, {Ferlet},
  {McConnell}, {Mayor}, \& {Parkinson}}]{VidalMadjar2004}
{Vidal-Madjar}, A., {D{\'e}sert}, J.-M., {Lecavelier des Etangs}, A., {et~al.}
  2004, \apjl, 604, L69

\bibitem[{{Walkowicz} \& {Basri}(2013)}]{WalkowiczB13}
{Walkowicz}, L.~M., \& {Basri}, G.~S. 2013, \mnras, 436, 1883

\bibitem[{{Winn} {et~al.}(2010){Winn}, {Fabrycky}, {Albrecht}, \&
  {Johnson}}]{WinnFAJ10}
{Winn}, J.~N., {Fabrycky}, D., {Albrecht}, S., \& {Johnson}, J.~A. 2010, \apjl,
  718, L145

\bibitem[{{Wu} \& {Lithwick}(2011)}]{WuLithwick11}
{Wu}, Y., \& {Lithwick}, Y. 2011, \apj, 735, 109

\bibitem[{{Wu} \& {Murray}(2003)}]{WuMurray03}
{Wu}, Y., \& {Murray}, N. 2003, \apj, 589, 605

\bibitem[{{Zahn}(1966)}]{Zahn1966}
{Zahn}, J.~P. 1966, Annales d'Astrophysique, 29, 489

\bibitem[{{Zahn}(1977)}]{Zahn1977}
{Zahn}, J.-P. 1977, \aap, 57, 383

\bibitem[{{Zahn}(1989)}]{Zahn1989}
---. 1989, \aap, 220, 112

\bibitem[{{Zhang} \& {Penev}(2014)}]{ZhangPenev14}
{Zhang}, M., \& {Penev}, K. 2014, ArXiv e-prints, arXiv:1404.4365

\end{thebibliography}
\bibliographystyle{apj}

\end{document}